Prediction of polymer mixture compatibility by Monte Carlo simulation of intermolecular binary interactions

Amirhossein Ahmadi and Juan J. Freire*

Departamento de Ciencias y Técnicas Fisicoquímicas, Facultad de Ciencias,
Universidad Nacional de Educación a Distancia, 28040 Madrid, Spain.

Running Head: Intermolecular interactions in polymer mixtures

* Corresponding author. Phone/fax: 34-913988627. e-mail: jfreire@invi.uned.es

ABSTRACT

We have evaluated conformational and orientational averages of binary interaction integrals for pairs of chains constituting atomistic representations of short polymer molecules. By considering A-A, B-B and A-B pairs, we relate these results with the Flory-Huggins parameter for the A-B mixtures. This parameter is commonly accepted as a good indicator of compatibility. Since the method ignores the simultaneous interactions with other molecules in the mixture, the local environment is approximately described by introducing an effective medium dielectric constant whose value is conveniently parameterized. The results for four different real systems are compared with data obtained from experimental neutron scattering data. The method qualitatively predicts the sign and variation with temperature in the four different cases, also showing a reasonable quantitative agreement in some of the cases. Its performance is discussed in comparison with a standard method that evaluates the Flory-Huggins parameter by calculating an average of the intermolecular energy of two molecules in contact, taking also into account their off-lattice Flory-Huggins coordination numbers.

# 1. Introduction

Knowing the degree of compatibility in polymer mixtures is of great scientific and technological interest, since its accurate prediction may lead to the design of new materials with better characteristics than the single components [1]. From the experimental point of view, compatibility of blends can be characterized by means of the Flory-Huggins parameter, $\chi$ [2]. This parameter can be experimentally extracted from data obtained in neutron scattering experiments. Although these experimental values of $\chi$ may not always completely consistent with the mixture separation diagrams, they provide a valuable indication on the mixture compatibility, which can theoretically be used for other systems containing different chain lengths.

Theoretical predictions of $\chi$ can be performed with numerical data obtained from numerical simulations [3]. Atomistic representations of the chains involved in the mixtures can be built and realistic interactions between different bonded or non-bonded atom can be incorporated from the information contained in "forcefield" files.

Molecular Dynamics simulations may actually be performed for simulation boxes representing the real systems. These simulations yield numerical trajectories of the polymer blends that constitute statistical samples from which different properties can be calculated. A collective "scattering" structure factor, $S(q)$, (similar to that used to describe the experimental scattering data) can be obtained from the trajectories. Its subsequent fit to a formula provided by the Random Phase Approximation provides a numerical estimations of $\chi$,

$$V_0 S^{-1}(q) = \frac{1}{N_A \Phi_A P_A(q)} + \frac{1}{N_B \Phi_B P_B(q)} - 2\chi \tag{1}$$

This procedure is similar to the treatment used for the experimental scattering data. In Eq. (1), $V_0$ is the reference microscopic volume, $V_R$, normalized with respect to the volume of a scattering unit, $N_i$ is the number of scattering units in chains $i$, and $\Phi_i$ is their volume fraction in the mixture. Finally, $P_i(q)$ is the individual scattering form factor of chain $i$, which can also be obtained from the trajectories. This method has been followed in some recent investigations for different mixtures of polymers [4-6]. However, it requires an important amount of computational time, since long equilibrated trajectories have to be generated in order to obtain sufficiently accurate values of $S(q)$.

A computationally efficient approach for the estimation of $\chi$ was proposed time ago. This approach is based in the direct evaluation of the intermolecular energy between a molecule of a given type, $i$, and another molecule, $j$, in contact with it, averaged over different orientations [7,8]. With this end, the two molecules $i$ and $j$ are randomly generated with a common center of masses position and the orientation of molecule $j$ is randomly changed and subsequently translated until its van der Waals surface not longer overlaps with molecule $i$. The intermolecular energy between the two molecules is then obtained. The process is repeated with different random orientations in order to obtain a distribution of intermolecular energies, $P_{ij}(E)$. The averaged intermolecular energy is calculated as

$$\bar{E}_{ij} = \frac{\int E P_{ij}(E) e^{\frac{-E}{RT}} dE}{\int P_{ij}(E) e^{\frac{-E}{RT}} dE} \tag{2}$$

According to the Flory-Huggins theory, a coordination number should be also

5determined. With this end, *j* molecules are subsequently introduced in a cluster with molecule *i*, following the same procedure described above, so that each new molecule *j* is randomly rotated and then translated to avoid any contact of its van der Waals surface with that of the cluster while there is a single contact of the new molecule with molecule *i*. The total number of *j* molecules introduced this way before not more molecules can be included determines the coordination number. The process is repeated many times to obtain the average coordination number $Z_{ij}$.

For a given mixture AB, parameter χ corresponding to the mean molecular volume of the units can be obtained as

$$\chi = \frac{1}{2}\left(Z_{AB}\bar{E}_{AB} + Z_{BA}\bar{E}_{BA} - Z_{AA}\bar{E}_{AA} - Z_{BB}\bar{E}_{BB}\right)V_0 / RT \qquad (2)$$

where $V_0$ is now referred to the mean volume of the molecules (obviously $\bar{E}_{AB} = \bar{E}_{BA}$).

Although this approach has been shown to qualitatively predict the miscibility behavior of some types of blends, we have verified that it cannot give accurate results for same types of molecular structures, in particular those containing remarkably anisotropic groups as phenyl rings.

In the present work, we explore another type of procedure consisting in determining binary interaction integrals between pairs of units. These integrals are orientational and conformational averages that take into account the three possible types of intermolecular interactions between the units (AA, BB or AB) at different distances. We have made use of a basic theoretical treatment to relate binary integrals with parameter χ. There are two substantial differences between our procedure and the approach explained in the preceding paragraph. 1) Our treatment takes into account interactions at all different distances between the molecular centers of masses. 2) However, it ignores the simultaneous presence of other molecules around the interacting pair. Actually, we take into account their mean-field effect on Coulomb interactions by introducing a parameterized medium dielectric constant. With the same purpose, we also explore other details in the long-range potential, in particular the cut-off for the intermolecular interactions, selecting the options that best reproduce the experimental data.

The results obtained with this new method are compared with existing experimental data and our own evaluations from the alternatives methods mentioned above. We explore four different types of mixtures: polystyrene/polyvinylmethylether, $PS-PVME$, polystyrene/polymethylmethacrylate, $PS-PMMA$, polystyrene/(1,4-cis)polyisoprene, $PS-PI$, and polyoxyethylene/polymethylmethacrylate, $POE-PMMA$. It is known that the $PS-PVME$ system is miscible at room temperature and shows a lower critical solution temperature, the long molecular weight $PS-PMMA$ and $PS-PI$ systems are not miscible at room temperatures, showing upper critical solution temperatures, and the $POE-PMMA$ system is miscible at room temperature. Precise values of the Flory-Huggins parameter $\chi$ for all these mixtures have been documented from neutron scattering data [9].

**2. Theoretical scheme**

Our procedure should be consistent with the basic specifications of the Flory-Huggins theory, though lattice restrictions do not apply to our atomistic chain models and, moreover, the concept of coordination number is not considered in our calculations. With this end, we divide a macroscopic system in $N$ sites. Each one of these sites may be in different configurations and interact with neighboring sites. All the sites should show a similar behavior in different regions of the system. Therefore, we assume that the extensive thermodynamic properties of the system are the sum of identical contributions from all sites, i.e., the sites behave as small macroscopic regions and the interactions between them can be described by a mean-field potential. Therefore, the intermolecular energy of the system, $U$, can be divided into $N$ equivalent terms, $U_i$,

$$U = (1/2)\sum_i \sum_j U_{ij} = \sum_i U_i \tag{3}$$

where

$$U_i = (1/2)\sum_{j \neq i} U_{ij} \tag{4}$$

Similarly, the average free energy corresponding to the intermolecular interactions of the system, $A^{inter}$, is divided into $N$ equivalent terms and each one of these terms is calculated from an independent site partition function,

$$A^{inter} = NA_i^{inter} = -k_B TN \ln\left(z_i^{inter}\right) \tag{5}$$

where $k_B$ is the Boltzmann factor and $T$ the temperature. The intermolecular site classical partition function, $z_i^{inter}$, can be obtained by integration of the probabilities associated to site energy $U_i$ over all configurations

$$z_i^{inter} = \left(1/V^{N-1}\right)\int \prod_{i \neq j}\left(1+\chi_{ij}\right)d\mathbf{R}_{ij} \tag{6}$$

where the terms $\chi_{ij}$ are defined as

$$\chi_{ij} = e^{-U_{ij}/2k_BT} - 1 \tag{7}$$

The integrals on the right term of Eq. (6) can be evaluated by assuming that all the $\chi_{ij}$ terms are equivalent, though each one acts on a different region of the space so that multiple integrals can be simply decomposed as products. With these assumptions we obtain

$$z_i^{\text{inter}} \cong \left(1/V^{N-1}\right)\left[\int\left(1+\chi_{ij}\right)d\mathbf{R}_{ij}\right]^{N-1} = \left[1+(N-1)I_{sites}/V+(N-1)(N-2)\left(I_{sites}\right)^2/2V^2+...\right] \tag{8}$$

where

$$I_{sites} = 4\pi < \int\left(e^{-U_{ij}(R_{ij})/2k_BT}-1\right)R_{ij}^2 dR_{ij} > \tag{9}$$

is defined as the binary interaction integral between sites. Therefore, for high values of $N$,

$$z_i^{\text{inter}} \cong \sum_{k=0}^{N}\frac{\left(V^{N-k-1}N^k\right)}{k!}\left[I_{\text{inter}}\right]^k = V^{N-1}e^{-(N/V)I_{bin}^{FH}} \tag{10}$$

and, finally, the average free energy of the system can be written as

$$A^{\text{inter}} = -k_BT\left(N^2/V\right)I_{sites} \tag{11}$$

It should be commented that the total intermolecular energy can be obtained through an alternative mathematical procedure without assuming independent site contributions. In this case, the $N(N-1)/2$ interactions should be simultaneously included in the evaluation of the system partition function, $Z$, from the total intermolecular energy, $U$. Assuming that the binary interactions occur in different independent regions, the final result is

$$A^{\text{inter}} = -k_BT\left(N^2/2V\right)I_{bin} \tag{12}$$

This result includes the more standard binary interaction integral, $I_{bin}$,

$$I_{bin} = 4\pi < \int \left( e^{-U_{ij}(R_{ij})/k_B T} - 1 \right) R_{ij}^2 dR_{ij} > \qquad (13)$$

that also provides the well known relationship between second virial coefficients and intermolecular potentials. (This procedure is aimed to the study of dilute systems. In previous work, we have numerically predicted second virial coefficients for n-alkanes and branched-alkanes in the gas phase [10,11] and also for linear and star polymer solutions [12,13], through the evaluation of $I_{bin}$ for these chain molecules, using bead repeat units that interact through a Lennard-Jones potential.) Eq. (11) and Eq. (12) give similar results for the total intermolecular energy when the terms $U_{ij}(R)$ are largely positive at small $R$ and close to zero in the attractive region.

Eq. (11) can be easily generalized to the case of a mixture of A and B sites. The intermolecular energy of $N_A$ sites, each one interacting with $N_B$ sites is

$$A_{AB}^{\text{inter}} = -k_B T N_A (N_B/V) I_{sites}^{AB} \qquad (14)$$

where

$$I_{sites}^{AB} = \int_0^\infty f_M^{AB} dR_{ij} \qquad (15)$$

$f_M^{AB}$ is the averaged radial Mayer function of the binary interaction between site of A and B types,

$$f_M^{AB} = 4\pi R_{ij}^2 < \left( e^{-U_{ij}^{AB}(R_{ij})/2k_B T} - 1 \right) > \qquad (16)$$

and $I_{sites}^{AB}$ denotes the corresponding site binary interaction integral. It includes orientational and conformational averages over a sample of pair of sites A and B (chains or chain fragments) whose centers of masses are placed at a given distance $R_{ij}$. The mixture total intermolecular free energy is the balance of the 4 different types of interactions between the 2 different types of sites. Therefore, for a system composed of $N_A \Phi_A$ A-sites and $N_B \Phi_B$ B-sites,

$$A_{mixture}^{inter} = -k_B T \Phi_A \Phi_B \left[ 2(N_A N_B / V) I_{sites}^{AB} - (N_A^2 / V) I_{sites}^{AA} - (N_B^2 / V) I_{sites}^{BB} \right] \quad (17)$$

The density of sites ($N_i/V_i$) can be simply estimated from the molar volumes of sites A and B, calculated from their molecular weights and macroscopic densities, $(N_i / V) = N_{AV} / v_i$, where $N_{AV}$ is the Avogadro number and $v_i$ is the site molar volume. Experimental values of $\chi$ are usually given for a given reference site volume, $V_R$, common to different systems. The number of sites consistent with this definition is $N_i^R = V_R (N_i / V)$. Taking into account these terms, the comparison of Eq. (16) with the enthalpic term of the Flory-Huggins free energy, $A_{enthalpic}^{FH} = k_B T \Phi_A \Phi_B \chi$, finally yields

$$\chi = -V_R N_{AV}^2 \left[ 2 v_A^{-1} v_B^{-1} I_{sites}^{AB} - v_A^{-2} I_{sites}^{AA} - v_B^{-2} I_{sites}^{BB} \right] \quad (18)$$

## 3. Computational methods

We compute integrals $I_{sites}^{AB}$, $I_{sites}^{AA}$ and $I_{sites}^{BB}$ for the A and B components of the different mixtures. The interaction sites are defined to be short chains of similar sizes For *PS*, *PI* and *PMMA*, we use chains of 3 repeat units. For *PVME* and *POE*, we use chains with 5 repeat units and 4 repeat units, respectively. (The average radii of gyration of all these chains are in the range 3.7±0.2 Å).

Different methods can be employed to build chains and sample conformations. In our case it has been expedient to use the "Build" and "Discover" modules of the software suite "Materials Studio" [8]. We have picked up pairs of conformations contained in Molecular Dynamics simulation trajectories. These simulations have been performed on *NVT* simulation boxes of the pure components and mixture systems, with periodic boundary conditions. Temperature has been maintained approximately constant by means of the Andersen numerical thermostat [14]. The simulation boxes include 20 chains. Different configurations are obtained through relatively short simulation runs from previously equilibrated systems (1000 ps, with time steps of 1 fs). We have analysed 20 configurations or frames along the trajectories. For each one of these configurations, we have considered the interactions between the different 190 pairs of chain conformations, defined from the coordinates of all atoms inside the box at the different selected frames.

Our Monte Carlo procedure to sample orientations for a given pairs of chain conformations have been described in detail before [10]. We locate the centre of mass of one of the chains, A, in the origin of coordinates. The center of masses of the other chain, B, is placed according a fixed radial coordinate $R_{ij}$ and two randomly chosen polar angles. We also choose three additional Euler angles defining the orientation of the internal coordinate axes of chain B. This way, all the coordinates of atoms A and B are defined with respect to the coordinate axes of chain A. We repeat this process 1000 times for each given pair of conformations at each given distance $R_{ij}$. We compute the terms $e^{-U_{ij}^{AB}(R_{ij})/2k_BT}$, obtain the conformational and orientational averages and evaluate the numerical integral defined in Eq. (15).

The terms $U_{ij}^{AB}(R_{ij})$ can be obtained as a sum of all the different intermolecular interactions between atoms. These interactions are defined in forcefield files. van der Waals and Coulomb interactions should be included. In this work, we have employed the forcefield PCFF, based in ab initio calculations and specifically aimed to polymer chains [15].

Our theoretical treatment neglects simultaneous interactions between the sites (chains). However, other chains are actually located in the same space region where the pairs of chains A and B interact. We assume that these other chains can be treated as a macroscopic medium which can be approximately taken into account by properly modifying the interactions. In particular, the vacuum dielectric constant, $\varepsilon_0$ is substituted by a medium dielectric constant, $\varepsilon$. Distance-dependent dielectric constants are particularly useful to describe an implicit medium since it can give a more precise description of the stronger interactions at short distances [8,16-17]. We consider

$$\varepsilon = \varepsilon_0 ar \tag{18}$$

where $a$ is an empirical parameter, that we fix to give the most adequate reproduction of the experimental data. After analyzing our numerical results for the different systems, we have adopted the value $a=1.3$.

Long-range interaction potentials usually include a cut-off. In the case of chains in an implicit medium, attractive interactions at high distances are overestimated unless the cut-off value is reduced [18]. For the present case, we have verified that a slight reduction of the cut-off to the value of 8 Å, applied over a spline range of 1.1 Å, gives the most adequate numerical results. It should be remarked that the final results provided by our method are particularly sensitive to these details in the application of the interaction potential.

Our results are compared with calculations performed with the method based in the direct computation of intermolecular energies and coordination numbers, described in the Introduction, according to its implementation in module "Blends" of "Materials Studio". In this case, this method shows an even greater sensibility to the forcefield details. The most reasonable results have been obtained with the COMPASS forcefield without any potential cut-off. (This forcefield was provided to describe interactions of polymers in condensed state [19]. It should be considered that these calculations are performed for a pair or a cluster of chains in contact; therefore the prescriptions for condensed systems are more likely to apply).

We will also discuss some previous results [6] that we obtained by computing collective scattering functions for long Molecular Dynamics trajectories (60-84 ns) of the *PS-PVME* mixture. The latter simulations were performed in *NVT* simulation boxes with a reduced density of 0.7 g./cm$^3$. We computed Coulomb interactions according to the vacuum dielectric constant and selected the COMPASS forcefield with the standard cut-off value of 9.5 Å [8].

## 4. Results and discussion

In Fig. 1, we show numerical values for the averaged radial Mayer function defined in Eq. (16). These results are obtained through our Monte Carlo simulation for the *PS-PS*, *PS-PI* and *PI-PI* pairs of chains at 400 K. There is a purely repulsive region for which the energy is large and positive and the function is close to $-4\pi R^2$, followed by a significant attraction region, before reaching the asymptotic limit corresponding to chains too far away to interact.

Fig. 2 shows numerical values of the three different site binary interaction integrals, in the $PS-PI$ system at 400 K, as defined in Eq. (15). The plotted functions have been obtained with different upper integration limits for $R$, $R_{max}$. It is observed that practically constant integral values are obtained for $R>20$Å. The curve corresponding to the mixed pair (type AB) is between those corresponding to pairs of identical chains (AA or BB). Similar curves were obtained for all polymer blends and temperatures. Moreover, the AB integrals for the AB pairs are always intermediate between those corresponding to the AA and BB pairs.

Final results for $\chi$ obtained from Eq. (18) for different systems and temperatures are shown in Table 1. These results show a general qualitative agreement with the experimental estimations, obtained from neutron scattering data [9], also contained in Table 1. Thus, the simulation values for the Flory-Huggins parameter corresponding to the $POE-PMMA$ mixtures are slightly negative, i.e., these systems behave as miscible in agreement with experimental behavior. Our $\chi$ results are clearly positive for the $PS-PI$ mixtures, also in agreement with the experiments. For the $PS-PMMA$ mixtures, both simulation and experimental data are positive. Finally, our results for the $PS-PVME$ mixtures are small and change from negative to positive values as temperature increases, a feature also shown by with the experimental data. From the quantitative point of view, the simulation and experimental results are reasonably close for most systems, though some differences are important due to the small absolute value of the experimental data, clearly exceeding the uncertainty range of the simulation results. (The error bars of the simulation data are estimated to be small than ±0.01 through the analysis of deviations of several independent simulation runs). These differences are very significant in the case of the $PS-PMMA$ systems.

It can also be observed that, in spite of the quantitative discrepancies, both the calculations and the experimental data for $PS-PMMA$ and $PS-PI$ systems show temperature variations that are consistent with upper critical solution temperatures, while the $PS-PVME$ mixture data predict a lower critical solution temperature behavior. Although the χ data reported from the neutron scattering experiments for $POE-PMMA$ mixtures [9] do not vary with temperature, some phase separation studies [20] claim the existence of an upper critical solution temperature. Our simulation results indicate a small negative variation of χ with temperature, consistent with an upper critical solution temperature, though they indicate that this temperature would be placed below the mixture glass transition.

The results obtained with the method incorporated in module "Blends" (direct calculation of interaction energies and coordination numbers) are also reported in Table 1. In comparison with procedure proposed in this work, this method offers similar qualitative description of the experimental behavior for the $PS-PI$, $POE-PMMA$ and $PS-PMMA$ mixtures. From the quantitative point of view, the "Blends" module is in relatively closer agreement with the $PS-PMMA$ experimental data, gives results higher than our method for the $PS-PI$ mixtures and its performance is clearly worse, giving significantly high absolute values, for the $POE-PMMA$ blends.

Moreover, the "Blends" module offers a poor qualitative description of the experimental data for the $PS-PVME$ systems, showing high positive values of χ and predicting an upper critical solution temperature. Both features are in contradiction with experimental behavior. This discrepancy may be due to the method failure to cope with molecules containing anisotropic phenyl rings with important mutual attractive interactions.

Finally, we should compare the results in Table 1 for the $PS-PVME$ with our Molecular Dynamics estimations of $\chi$, based in fits of collective scattering structure functions obtained from the simulations to Eq. (1). In these previous work [6], the mean results for $\chi$ were reported to be -0.02 for the systems at 350 and 400 K and -0.006 for 450 K. It can be observed that, in comparison with the experimental data, these data from a more direct procedure have a performance similar to the simulation values obtained in this work using considerably less extensive calculations.

It is, therefore, concluded that the calculation of binary interaction integrals has lead to reasonable predictions of the miscibility behavior of different systems. However, this method requires the previous parameterization of some simulation parameters to take approximately into account many-chain effects that are not incorporated into the theoretical description.

**Acknowledgment** This work has been partially supported by Grant CTQ2006-06446 from DGI-MEC, Spain. AA also thanks Fellowships for Graduate Studies from the MAE, Spain and the Universidad Nacional de Educación a Distancia.

Table 1. χ (for a volume of 100 Å$^3$) obtained from the methods described in the text for different mixtures at different temperatures, compared with values computed from fitted curves that summarize neutron scattering data [9].

From Eq. (18)

| Temperature, K | PS-PVME | PS-PMMA | POE-PMMA | PS-PI |
|---|---|---|---|---|
| 298 | -0.0018 | 0.6 |  | 0.13 |
| 350 | -0.0006 | 0.26 | -0.011 | 0.057 |
| 400 | 0.006 | 0.15 | -0.016 | 0.023 |
| 450 |  |  | -0.019 |  |

Experimental

| Temperature, K | PS-PVME | PS-PMMA | POE-PMMA | PS-PI |
|---|---|---|---|---|
| 298 | -0.041 | 0.0195 | -0.0021 | 0.067 |
| 350 | -0.02 | 0.0187 | -0.0021 | 0.0581 |
| 400 | -0.0045 | 0.0178 | -0.0021 | 0.052 |
| 450 | 0.007 | 0.0172 | -0.0021 | 0.047 |

"Blends" module

| Temperature, K | PS-PVME | PS-PMMA | POE-PMMA | PS-PI |
|---|---|---|---|---|
| 298 | 0.32 | 0.12 | -0.67 | 0.21 |
| 350 | 0.24 | 0.09 | -0.27 | 0.19 |
| 400 | 0.19 | 0.09 | -0.04 | 0.12 |
| 450 | 0.16 | 0.07 | -0.02 | 0.12 |

**Figure captions**

**Fig. 1**. Numerical values of the averaged radial Mayer function defined in Eq. (16), for the different pairs of chains in the $PS-PI$ mixture at 400 K.

**Fig. 2**. Results for the site binary interaction integrals defined in Eq. (15), corresponding to the different site interactions in the $PS-PI$ mixture at 400 K, obtained with different numerical values of the upper integration limit, $R_{max}$.

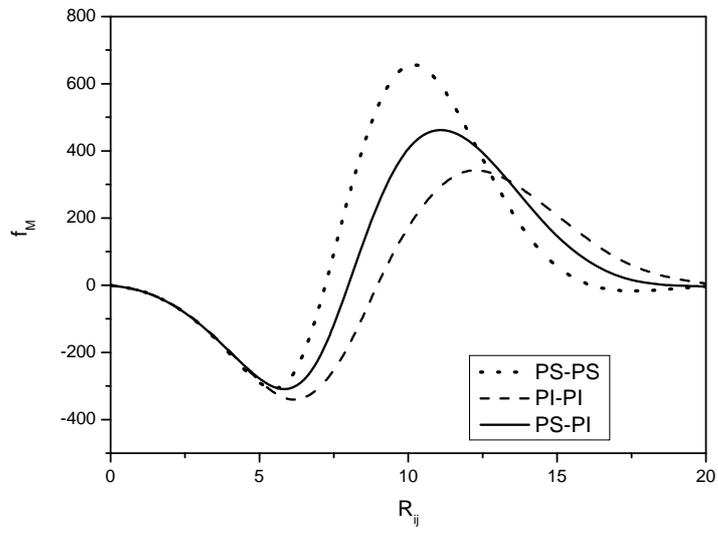

Figure 1

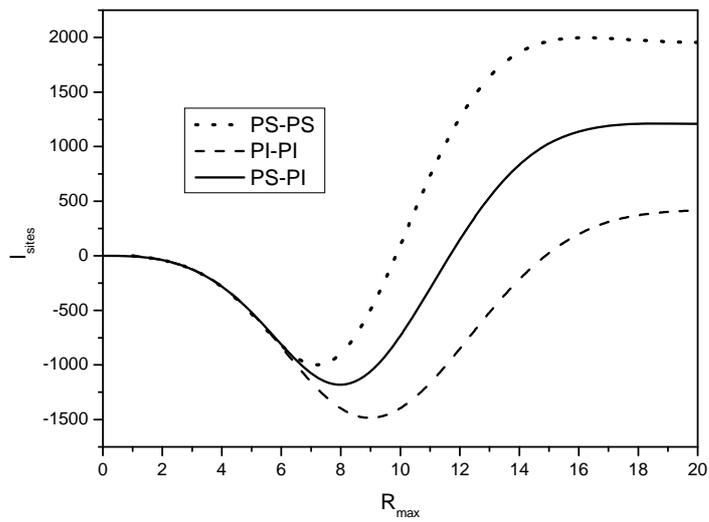

Figure 2